\documentclass[aip,rsi,reprint,superscriptaddress]{revtex4-1}
\usepackage{amssymb,amsmath}
\usepackage{graphicx,color}

\draft 

\begin{document}

\title{Wide-band current preamplifier for conductance measurements with large input capacitance}

\author{Andrey V. Kretinin}
\affiliation{Weizmann Institute of Science, Condensed Matter Physics Department, Rehovot, Israel}
\author{Yunchul Chung}
\email{ycchung@pusan.ac.kr}
\affiliation{Department of Physics, Pusan National University, Busan, 609-735, Republic of Korea}

\date{\today}

\begin{abstract}
A wide-band current preamplifier based on a composite operational amplifier is proposed. It has been shown that the bandwidth of the preamplifier can be significantly increased by enhancing the effective open-loop gain of the composite preamplifier. The described preamplifier with current gain 10$^7$~V/A showed the bandwidth of about 100~kHz with 1~nF input shunt capacitance. The current noise of the amplifier was measured to be about 46~fA/$\sqrt{\rm Hz}$ at 1~kHz, close to the design noise minimum. The voltage noise was found to be about 2.9~nV/$\sqrt{\rm Hz}$ at 1~kHz, which is in a good agreement with the value expected for the operational amplifier used in the input stage. By analysing the total noise produced by the preamplifier we found the optimal frequency range suitable for the fast lock-in measurements to be from 1~kHz to 2~kHz. To get the same signal-to-noise ratio, the reported preamplifier requires roughly 10$\%$ of the integration time used in measurements made with a conventional preamplifier.
\end{abstract}


\maketitle 

\section{Introduction}\label{Intro}

Current or transimpedance preamplifier is one of the most commonly used tools for signal acquisition from high-impedance devices under test (DUT) such as photodiodes and bolometers. It is also widely used to measure the low-frequency (10~Hz -- 100~Hz) conductance of various mesoscopic quantum devices, such as quantum dots,\cite{Meirav1990} quantum point contacts,\cite{Wees1988-Wharam1988} and electron interferometers\cite{Yacoby1996} by detecting the device current in the constant voltage regime using the lock-in technique. The overwhelming majority of the mesoscopic quantum devices operate at cryogenic temperatures (below 4.2~K), as a consequence the cryogenic DUT has to be connected to the room-temperature preamplifier by means of long signal lines of the cryostat. Apart from that, the signal lines are often heavily filtered to suppress the external RF noise. Combination of the stray capacitance of the lines and filters capacitance results in a relatively large capacitance (0.5~nF -- 5~nF) shunting the input of the preamplifier. This large shunting capacitance reduces the measurement bandwidth of the preamplifier and increases the overall background noise, which in turn reduces the signal-to-noise ration (SNR) and requires longer data acquisition time. Therefore, it is important to develop a transimpedance preamplifier, which would provide a large (DC to $\sim$100~kHz) bandwidth with a large ($\sim$1~nF) input capacitance.

Recently, Vandersypen \emph{et al.} developed a current preamplifier based on a dual low-noise junction field effect transistor (JFET) with the open-loop gain of about 10$^4$, and demonstrated the bandwidth exceeding 100~kHz at 1.5~nF with the current gain of $10^7$~V/A.\cite{Vandersypen2004} However, it is cumbersome to build a large open-loop op-amp using discrete components, moreover the authors did not provide an adequate description of the op-amp circuit itself. Usually, it is not so difficult to find an op-amp with high open-loop gain based on bipolar junction transistors (BJT), but they produce too high input current noise (few pA/$\sqrt{\rm Hz}$) to be used for a current amplifier. Meanwhile, op-amps with JFET input stage have rather low current noise (at least three orders of magnitude smaller than that for BJT), but small open-loop gain. To overcome these complications, we used a so-called 'composite op-amp',\cite{Mikhael1987-KalthoffAppBull} which is a cascaded low noise JFET op-amp input stage and a wide-band current feedback op-amp. This realization makes it much easier to build a wide-band current amplifier with low input noise.

In this report we present a simple transimpedance preamplifier based on a composite op-amp design, which was built using the BiFET AD743 op-amp as the low-noise input stage, and the high performance AD811 op-amp as the gain booster. The increased open-loop gain allowed us to extend the bandwidth up to 100~kHz at 1~nF input capacitance with the current gain of $10^7$ V/A. The total current and voltage noise of the preamplifier measured at 1~kHz ware 46~fA/$\sqrt{\rm Hz}$ and 2.9~nV/$\sqrt{\rm Hz}$ correspondingly, which is comparable to that reported for the current preamplifier built from discrete JFETs.\cite{Vandersypen2004}

\section{The bandwidth of the current preamplifier}\label{Bandwidth}

\begin{figure}
  \includegraphics[width = \columnwidth]{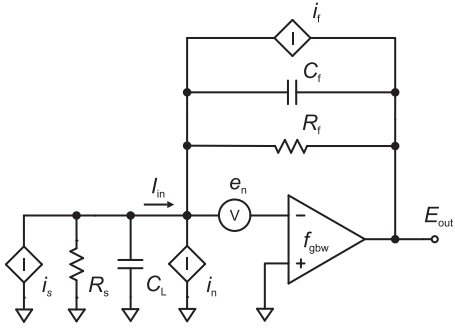}\\
  \caption{The full equivalent circuit of the transimpedance inverting preamplifier formed by the operational amplifier with gain bandwidth $f_{\rm gbw}$ and feedback network $R_{\rm f}C_{\rm f}$. The device under test is modeled with resistor $R_{\rm s}$ connected by the signal line with capacitance $C_{\rm L}$.}
  \label{Fig1}
\end{figure}

First we would like to discuss the factors determining the bandwidth of an op-amp based transimpedance preamplifier. Figure~\ref{Fig1} shows the equivalent circuit of the preamplifier formed by an op-amp with the gain bandwidth $f_{\rm gbw}$, and the feedback network $R_{\rm f}C_{\rm f}$. The DUT is represented by resistor $R_{\rm s}$ shunted with input capacitance $C_{\rm L}$. The meaning of the current and voltage sources will be discussed later in Sec.~\ref{Noise}.

The transimpedance preamplifier is prone to be unstable, especially with large input capacitance, hence usually requires some feedback compensation with $C_{\rm f}$. The preamplifier transfer function is $V_{\rm out} = -I_{\rm in}\cdot R_{\rm f}\left[1+(\omega R_{\rm f}C_{\rm f})^2\right]^{-1/2}$, where $\omega = 2\pi f$, and it defines the high-frequency cut-off as
\begin{equation}
f_{\rm c} = \frac{1}{2\pi R_{\rm f}C_{\rm f}}.
\label{f_c}
\end{equation}
The chose of the value of $C_{\rm f}$ is determined by the stability considerations, and strongly affected by the value of $C_{\rm L}$ coupled to the input. Indeed, for the circuit in Fig.~\ref{Fig1} to be stable the feedback zero frequency $f_{\rm z} = (2\pi (R_{\rm s}\parallel R_{\rm f})(C_{\rm L}+C_{\rm f}))^{-1}$ should be inside the open-loop gain curve determined as $A = f_{\rm gbw}/f$.\cite{Wang2005} The largest stable bandwidth in this case is set by the condition $A\cdot\beta = 1$ at $f = f_{\rm c}$, where $\beta$ is the feedback factor.\cite{footnote1} This condition is used to find an optimal value of the feedback capacitance, under condition $R_{\rm s}C_{\rm L}\ll R_{\rm f}C_{\rm f}$,
\begin{equation}
C_{\rm f} = \sqrt{\frac{C_{\rm L}}{2\sqrt{2}\pi f_{\rm gbw}R_{\rm f}}}.
\label{C_f}
\end{equation}
As seen from the Eq.~(\ref{f_c}) and Eq.~(\ref{C_f}) the input capacitance $C_{\rm L}$ reduces the amplifier bandwidth, and a high open-loop gain op-amp is needed to compensate for this reduction. Alternatively, one can reduce the value of $R_{\rm f}$, but in this case SNR will be seriously compromised since it is proportional to $\sqrt{R_{\rm f}}$. Assuming the input signal is $I_{\rm in}$, and the dominant background noise is dictated by the thermal noise of $R_{\rm f}$ as $I_{\rm noise} = \sqrt{4k_{\rm B}T/R_{\rm f}}$, which results in $\rm{SNR} = I_{i\rm n}/I_{\rm noise} \propto \sqrt{R_{\rm f}}$.

Another limitation for the bandwidth of the transimpedance preamplifier comes from the fact that the op-amp gain is finite, and the shunting capacitance $C_{\rm L}$ affects the transfer function. The naive explanation would be that at higher frequencies the input current $I_{\rm in}$ is shunted by $C_{\rm L}$, and part of it is diverted to the ground, bypassing the effective input impedance of the current preamplifier $R_{\rm in}=R_{\rm f}/(1+A)$.\cite{footnote2}. In this case the -3dB cut-off frequency is
\begin{equation}\label{f_in}
    f_{\rm -3~dB} = \sqrt{\frac{(\sqrt{2}-1)f_{\rm gbw}}{2\pi R_{\rm f}C_{\rm L}}} = f_{\rm c}\cdot\sqrt{(\sqrt{2}-1)\cdot\frac{C_{\rm f}}{C_{\rm L}}\cdot\frac{f_{\rm gbw}}{f_{\rm c}}}.
\end{equation}
Note, that in any reasonable situation $C_{\rm L}\gg C_{\rm f}$ and $f_{\rm gbw}\gg f_{\rm c}$, hence $f_{\rm -3~dB}<f_{\rm c}$ and the real-situation bandwidth is determined by Eq.~(\ref{f_in}) rather than by Eq.~(\ref{f_c}), unless the feedback network is overcompensated ($C_{\rm f}$ is larger than the optimal value given by Eq.~(\ref{C_f})). It is clear from Eq.~(\ref{f_c}) and Eq.~(\ref{f_in}) that the only way to compensate the bandwidth reduced by $C_{\rm L}$ is to increase the open-loop gain of the op-amp.

\section{Composite OP-AMP current amplifier}\label{Design}

\begin{figure}
  \includegraphics[width=\columnwidth]{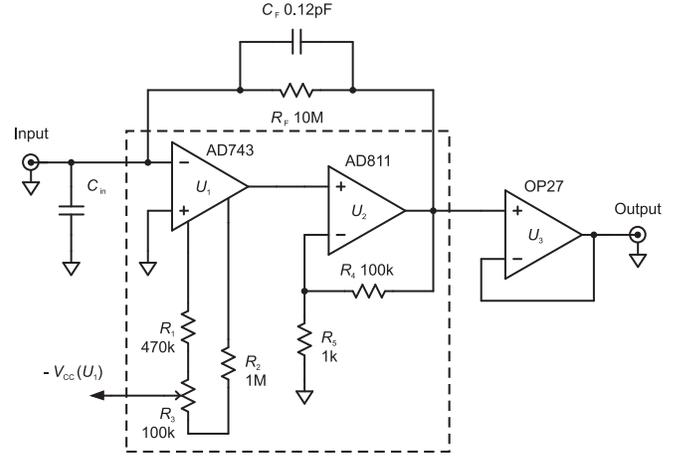}\\
  \caption{The circuit diagram of the composite current preamplifier with the current gain 10$^7$~V/A. The composite op-amp part of the circuit is marked by the dashed box.}
  \label{Fig2}
\end{figure}

Most of the commercial JFET input op-amps, which are suitable for the current preamplifier application, have a relatively small gain bandwidth. For example, an ultralow-noise high speed AD743 op-amp\cite{AD743DataSheet} has excellent input noise characteristics, but with the rather low gain bandwidth $f_{\rm gbw}=$~4.5~MHz. Therefore, it is difficult to use it for a wide-band applications, since the preamplifier bandwidth with a 1~nF shunt capacitor and $R_{\rm f}=$~10$^7$ would be about 10~KHz. To overcome this problem we adopted the so-called composite op-amp design.\cite{Mikhael1987-KalthoffAppBull} The circuit diagram of our composite amplifier is shown in Fig.~\ref{Fig2}. As an input stage we used the ultra-low noise AD743 op-amp, $U_{1}$. The input stage is followed by the gain booster built as a non-inverting voltage amplifier on the current feedback op-amp AD811, $U_{2}$, with the voltage gain of 10$^2$. Both stages share the same feedback network $R_{\rm f}C_{\rm f}$, and form the composite op-amp (enclosed in the dashed box in Fig.~\ref{Fig2}) featured with the low-noise input and the gain bandwidth enhanced by 10$^2$ ($f_{\rm gbw}=$~450~MHz).
The preamplifier is completed with the voltage buffer, $U_{3}$.

\begin{figure}
  \includegraphics{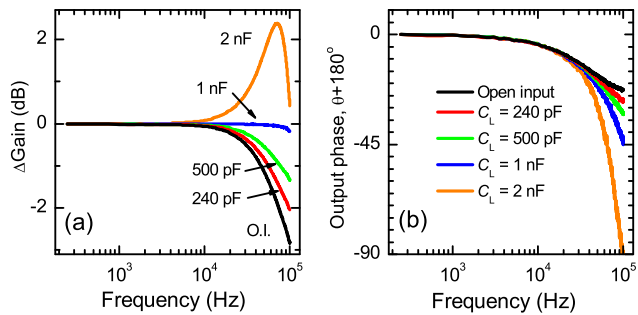}\\
  \caption{(Color online) The frequency response of the composite current preamplifier shown in Fig.~\ref{Fig2}. (a) The gain flatness at different values of input capacitance $C_{\rm L}$ illustrated by the normalized gain $\Delta\rm{Gain} = 20\log{(-V_{\rm out}/I_{\rm in}R_{\rm f})}$. (b) The signal phase frequency response at different values of the input capacitance $C_{\rm L}$. The value of $C_{\rm L}$ increases for curves from bottom to top.}
  \label{Fig3}
\end{figure}

Several op-amps were tested as the gain booster, and the current-feedback video op-amps usually gave better high frequency characteristics. However, setting the booster gain higher than 10$^2$ usually caused the circuit instability. The input offset voltage of the composite current preamplifier can be adjusted by nulling the first stage op-amp with a conventional trimming circuit $R_{1}\div R_{3}$ ($R_{3}$ is the multi-turn wire-wound type). To reduce the stray capacitance of the feedback resistor, $R_{\rm f}$ was made of five 2~MOhm thick-film resistors connected in series. The value of $C_{\rm f}$ was found experimentally by stabilising the preamplifier with the 1~nF capacitor connected to the input. The nominal value of $C_{\rm f}=$~0.12~pF is close to that found from Eq.~(\ref{C_f}) ($\sim$0.15~pF), and it was achieved by series connection of four 0.47~pF surface mount ceramic capacitors. The open-input capacitance $C_{\rm in}$ was estimated to be about 47~pF.

Apart from a pair of ceramic and electrolyte decoupling capacitors (47~nF and 10~$\mu$F, correspondingly) connected to the power supply pins of every op-amp, the power lines were heavily filtered with $LC$ low-pass filters, or by the feedthrough $\pi$-sections mounted onto the preamplifier chassis. To decouple the low-frequency environmental including electric power noise, the preamplifier was powered by a pair of 12~V lead-acid batteries.

The frequency response of our composite current preamplifier for different values of the input capacitance $C_{\rm L}$ is presented in Fig.~\ref{Fig3}. Figure~\ref{Fig3}(a) illustrates the gain flatness given by the normilized gain $\Delta\rm{Gain} = 20\log{(-V_{\rm out}/I_{\rm in}R_{\rm f})}$. As seen from the plot, the open input gain (marked as `O.I.') has its cut-off frequency slightly above 100~kHz. The preamplifier gain has excellent flatness up to $f\approx$~10~kHz, and exhibits some peaking only when $C_{\rm L}$ exceeds 1~nF.

The gain overshoot masks the -3~dB cut-off point and it is not straight forward to determine the bandwidth of the amplifier. To do this we used the signal phase frequency response, and found the bandwidth as the frequency at which the signal phase shift is 45 degrees.\cite{HorowitzHillBook} From the frequency response shown in Fig.~\ref{Fig3} it is seen that the preamplifier bandwidth is around 100 kHz for 1~nF input capacitance, which is comparable to that of the current amplifier built using discrete JFETs.\cite{Vandersypen2004} Further improvement of the bandwidth can be achieved by using JFET op-amps with larger $f_{gbw}$, such as AD745 and AD8610. In our search for alternative op-amps we considered only those, which have the possibility to balance the input offset voltage.

\section{Noise characteristics}\label{Noise}

After the frequency response of the described preamplifier has been verified, we focus on the issue of noise performance. In the equivalent circuit of the preamplifier shown in Fig.~\ref{Fig1} all the components are assumed to be noiseless. The current and voltage fluctuations in the circuit are introduced by adding corresponding voltage and current noise sources. Here we consider the composite op-amp as a single entity, which is characterized by its equivalent input voltage ($e_{\rm n}$) and current ($i_{\rm n}$) noise.\cite{MotchenbacherBook} Another source of noise in the circuit is the thermal fluctuations of the feedback resistor $R_{\rm f}$ given by the current noise density $i_{f} = \sqrt{4k_{\rm B}T/R_{\rm f}}$. The presence of the DUT also contributes to the total noise through the thermal fluctuations of $R_{\rm s}$ with current density $i_{\rm s} = \sqrt{4k_{\rm B}T/R_{\rm s}}$. Assuming that all sources of noise are uncorrelated and the op-amp gain $A$ is large, the square of total output voltage noise density $V^2_{\rm n}$ can be written as
\begin{widetext}
\begin{equation}\label{Total_Noise}
    V^2_{\rm n} = \frac{R^2_{\rm f}}{1+(\omega R_{\rm f}C_{\rm f})^2}\cdot I^2_{\rm n} = \frac{R^2_{\rm f}}{1+(\omega R_{\rm f}C_{\rm f})^2}\left[i^2_{\rm s}+i^2_{\rm f}+i^2_{\rm n}+e^2_{\rm n}\left(\frac{1}{R_{\rm s}}+\frac{1}{R_{\rm f}}\right)^2+e^2_{\rm n}\left(\omega (C_{\rm L}+C_{\rm f})\right)^2\right].
\end{equation}
\end{widetext}
Here $I_{\rm n}$ is the total input current noise density, and the prefactor $R^2_{\rm f}\left[1+(\omega R_{\rm f}C_{\rm f})^2\right]^{-2}$ accounts for the finite bandwidth of the preamplifier.

To characterise the the preamplifier noise, and quantify the contribution of each of the noise sources, we measured the thermal noise spectral density of several different metal film resistors at $T=$~4.2~K. These resistors were placed inside the liquid He$^4$ dewar on the experimental probe. Each of the resistor was connected by a coaxial cable producing the total shunt capacitance $C_{\rm L}=$~438~pF. The values of resistor $R_{\rm s}$ were chosen to be close to the values typical for mesoscopic quantum devices ($R_{\rm s}\geq G^{-1}_{0}\approx$~12.9~kOhm, where $G_{0}=2e^2/h$ is the conductance quantum).\cite{Meirav1990,Wees1988-Wharam1988,Yacoby1996} The noise spectral density was measured with SR770 FFT Spectrum Analyzer (Stanford Research Systems). In order to increase SNR of the noise measurements, we used an additional low-noise voltage preamplifier (LI-75A, NF Corporation) in series with the tested current preamplifier.

\begin{figure}
  \includegraphics{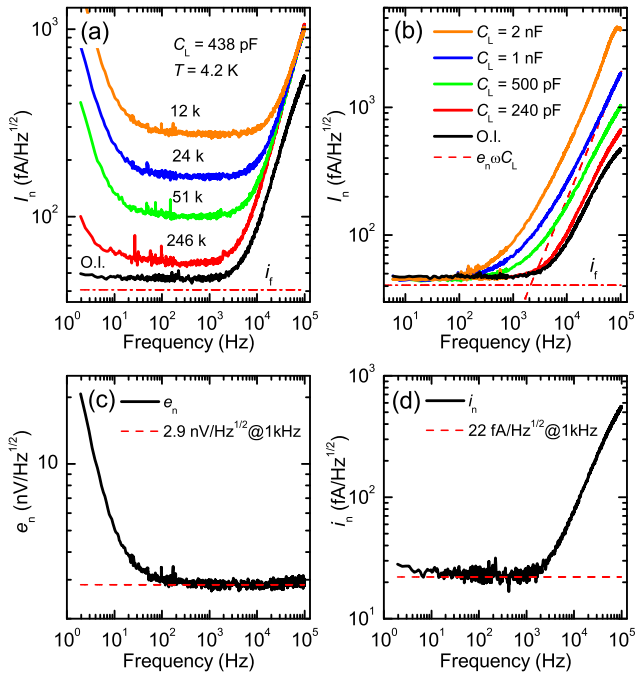}\\
  \caption{(Color online) The noise characterisation of the current preamplifier based on the composite op-amp shown in Fig.~\ref{Fig2}. (a) The total current preamplifier noise density spectrum measured for an open input (`O.I.') and different values of $R_{\rm s}$ (12~kOhm, 24~kOhm, 51~kOhm, 246~kOhm). $R_{\rm s}$ was immersed in liquid He$^4$ ($T=$~4.2~K) and connected by the coaxial cable with $C_{\rm L}=$~438~pF. The red dash dotted line represents the thermal noise of the feedback resistor $R_{\rm f}$ ($i_{\rm f}=$~41~fA/$\sqrt{\rm Hz}$). (b) The total current noise of the open-input preamplifier measured for different input capacitances $C_{\rm L}$. The dashed line shows the current noise expected from coupling of the voltage noise $e_{\rm n}=$~2.9~nV/$\sqrt{\rm Hz}$ through capacitor $C_{\rm L}=$~1~nF. The dash dotted line represents the thermal noise of the feedback resistor. (c) and (d) The extracted op-amp voltage and current noise density spectra, correspondingly.}
  \label{Fig4}
\end{figure}

The total current noise spectral density $I_{\rm n}$ measured for different $R_{\rm s}$ is shown in Fig.~\ref{Fig4}(a). On the same figure we plotted the current noise spectrum for the open input preamplifier (marked as `O.I.'). The open input noise represents the minimal possible noise of the current preamplifier, and it is estimated to be about 46~fA/$\sqrt{\rm Hz}$. This noise consists of the thermal noise of the feedback resistor $i_{\rm f}$, and the current noise of the composite op-amp $i_{\rm n}$  (contribution from the voltage noise $e_{\rm n}$ is negligibly small). As seen from Fig.~\ref{Fig1}(a), the thermal noise of $R_{\rm f}$ (41~fA/$\sqrt{\rm Hz}$), shown by the dash dotted line, is the dominant contributor and defines the absolute noise floor. The high-frequency increase observed at $f>$~2~kHz is typical for the current noise of an op-amp.\cite{MotchenbacherBook} As expected from Eq.~(\ref{Total_Noise}) the total noise increases if DUT and cable are connected to the preamplifier. The increase occurs due to the contribution from the thermal noise $i_{\rm s}$, and coupling of the voltage noise of the op-amp $e_{\rm n}$ through resistance $R_{\rm s}$ and capacitance $C_{\rm L}$ (last two terms in Eq.~(\ref{Total_Noise})). The thermal noise of the DUT is independent on frequency and adds equally to the whole spectrum. The voltage noise, on the contrary, has a $1/f$ component\cite{MotchenbacherBook} which is responsible for the noise increase at $f<$~100~Hz. The relatively frequency-independent minimum of the total noise at a given $R_{\rm s}$ occurs in the frequency range from 100~Hz to 2~kHz. This noise floor is mainly determined by contributions from $i_{\rm s}$ and $e_{\rm n}/R_{\rm s}$. At higher frequencies ($f>$~2~kHz) the voltage noise of the op-amp becomes also coupled through the input capacitance as $e_{\rm n}\omega(C_{\rm L}+C_{\rm f})$, and adds to the increasing high-frequency noise, initially brought by the op-amp current noise $i_{\rm n}$. Despite the fact that capacitor is noiseless, the input voltage noise coupled to the input capacitor introduces a significant share of the total noise, which is often overlooked. To illustrate how $e_{\rm n}$ is coupled through the input capacitance, we measured the total noise of the preamplifier with only $C_{\rm L}$ connected to the input. Figure~\ref{Fig4}(b) clearly shows that the high-frequency current noise increases with increasing capacitance $C_{\rm L}$ due to larger contribution from the op-amp voltage noise.

Assuming that the voltage source $e_{\rm n}$ does not contribute to the open-input noise, and the thermal noises of $R_{\rm f}$ and $R_{\rm s}$ are known, one can extract the spectral density of the op-amp noise. Figures~\ref{Fig4}(c) and \ref{Fig4}(d) show the spectrum of the voltage ($e_{\rm n}$) and current ($i_{\rm n}$) noise of the op-amp, correspondingly. As noticed before, $e_{\rm n}$ has a $1/f$ component for $f<$~100~Hz and becomes frequency independent at higher frequencies. The voltage noise density at 1~kHz is found to be about 2.9~nV/$\sqrt{\rm Hz}$, which is the value expected for the voltage noise of AD743.\cite{AD743DataSheet} The current noise $i_{\rm n}$ is frequency-independent up to 2~kHz with the minimum noise density of about 22~fA/$\sqrt{\rm Hz}$ at 1~kHz, which is about 3 times higher than that expected for the current noise of AD743. Most likely, the discrepancy originates from the unaccounted noise of the gain booster stage of the composite op-amp.

Figure~\ref{Fig4}(a) gives an idea on the optimal frequency range suitable for the lock-in measurements with this current preamplifier. It is clear that the lowest total noise occurs in the range from 1~kHz to 2~kHz. The advantage of higher lock-in reference frequency is that it allows to avoid the low-frequency $1/f$-noise, and reduce the lock-in integration time (provided SNR is high enough), which makes the measurements faster. The excellent frequency response and noise characteristics made possible to use the described current preamplifier in real mesoscopic transport experiments on one-dimensional conductors\cite{Kretinin2010} and quantum dots\cite{Kretinin2011} performed at submilli-Kelvin temperatures. The lock-in measurements using the composite preamplifier at $\sim$1~kHz with 30~ms integration time gave SNR comparable to that measured with the conventional current preamplifier (ITHACO 1211) at frequencies below 100~Hz and 300~ms integration time.

\section{Conclusion}\label{Concl}

A simple low-noise large bandwidth current preamplifier based on the composite operational amplifier is proposed. The composite operational amplifier has been employed to increase the open loop gain. The composite part was implemented by a low-noise BiFET input stage followed by a non-inverting high-speed voltage gain booster with common feedback network. The designed preamplifier had  current gain 10$^7$~V/A, and showed the bandwidth up to 100~kHz with 1~nF capacitor across the input. Apart from the large bandwidth, the described preamplifier demonstrated excellent noise characteristics. The current noise was found to be around 46~fA/$\sqrt{\rm Hz}$ at 1~kHz, which is close to absolute minimum defined by the thermal noise of the feedback resistor. The voltage noise was measured to be 2.9~nV/$\sqrt{\rm Hz}$ and shown to give a significant contribution to the total current noise when the DUT resistance and input capacitance are finite. The proposed current preamplifier was proven to be an excellent tool in mesoscopic quantum transport experiments.

\begin{acknowledgments}
Authors thank Moty Heiblum for making this work possible in Braun Center for Submicron Research. We also thank Vladimir Umansky, Diana Mahalu and Hadas Shtrikman for helpful discussion and experimental support. This work was partially supported by the National Research Foundation of Korea (NRF) grant funded by the Korea government (MEST) (No.2011-0003109) and the Korea Science and Engineering Foundation (KOSEF) grant funded by the Korea government (MEST) (No. 2010-0000268).
\end{acknowledgments}


\begin{thebibliography}{99}
\bibitem{footnote1} The feedback factor $\beta$ is the part of the output voltage applied to the input through the feedback network, and at $f=f_{\rm c}$ is $\beta = V_{\rm in}/V_{\rm out} = \frac{\sqrt{2}}{2}\left(1+\frac{R_{\rm f}}{R_{\rm s}}\right)\sqrt{1+(f_{\rm c}/f_{\rm z})^2}$, where $V_{\rm in}$ is the voltage at the inverting input.
\bibitem{footnote2} The transfer function in case of large $C_{\rm L}$ is $V_{\rm out} = -I_{\rm in}\cdot R_{\rm f}\left[\omega R_{\rm in}C_{\rm L}+\sqrt{1+(\omega R_{\rm f}C_{\rm f})^2}\right]^{-1}$.
\end{thebibliography}
\end{document}